\documentclass{ws-procs9x6-cpt16}
\begin{document}

\newcommand{\refeq}[1]{(\ref{#1})}
\def\etal {{\it et al.}}

\title{Slow Muons and Muonium}

\author{K.\ Kirch}

\address{Institute for Particle Physics, ETH Z\"urich\\
Otto-Stern-Weg 5, 8093 Z\"urich, Switzerland}

\address{Laboaratory for Particle Physics, Paul Scherrer Institut\\
5232 Villigen-PSI, Switzerland}

\begin{abstract}
The Paul Scherrer Institut in Switzerland operates the  high intensity proton accelerator facility HIPA.
A 590\,MeV kinetic energy proton beam of presently up to 2.4\,mA 
is sent 
to target stations producing pions, muons and neutrons for fundamental and applied physics. 
The  beam power of 1.4\,MW provides
the world's highest intensities of low momentum muons which can be stopped in low mass targets. Rates of surface muons 
of up to about $10^8$/s are being provided to various unique precision particle physics experiments.
Two feasibility studies are ongoing to considerably improve the available muon beams.
The high intensity muon beamline, HiMB, could deliver on the order of $10^{10}$/s surface muons 
and the stopped muon cooler, muCool, aims at a gain factor of $10^{10}$ 
in phase space quality while sacrificing only less than 3 orders of magnitude in intensity for low energy $\mu^+$. 
These beams will allow a new generation of precision physics experiments with stopped muons
and muonium atoms.
\end{abstract}

\bodymatter

\section{Muons brief and biased}
Muons have been undispensible probes in fundamental (and applied) physics for a long time. Their completely unexpected discovery is ascribed to Anderson and Neddermeyer.\cite{And36} Muons have already been seen earlier by Kunze\cite{Kun33} albeit without claiming detection of a new particle. Actually muons were until the 1940s mostly considered to be the predicted and searched for pions. In 1960 muonium (Mu), the hydrogen-like bound state of a positive muon and an electron ($\mu^+$e$^-$), was unambiguously detected by Hughes and coworkers.\cite{Hug60}
Muons and Mu continue to play a role of utmost importance in precision tests of the Standard Model of particle physics as well as in the search for new physics beyond it. They can be produced in comparatively large quantities, they live long enough for many applications and 
they decay sufficiently fast  in a parity violating, self polarization-analyzing weak process
($\mu^+ \rightarrow {\rm e}^+ \bar{\nu}_\mu \nu_{\rm e}$) which makes them most versatile. 
Actually, the discovery of parity violation (PV) in nuclear $\beta$ decay\cite{Wu57} was instantaneously followed by the demonstration of PV in muon decay.\cite{Gar57, Fri57} Later it has been shown that the V-A structure of the Standard Model weak interaction follows already solely from muon decay experiments, including inverse muon decay.\cite{Fet86}
The reader will find an excellent review on fundamental muon physics\cite{Gor15} and many references concerning the important role of low momentum muons, pions and neutrons in precision physics in Ref.\ \refcite{Kir15}. 
Below, a brief status update on muon activities at PSI is presented without explicitly connecting to the Standard-Model Extension (SME) and its coefficients.\cite{Kos16} However, given the fact that all precision experiments acquire time stamped data sets, a wealth of opportunities exists for corresponding analyses.

\section{Recent fundamental muon physics at PSI}
PSI's HIPA complex with its unique beam power provides the largest intensities of low momentum pions, muons and ultracold neutrons to fundamental physics experiments. The arrangement of the pion production targets as part of the beam optics of the proton beam onto the continuous spallation neutron source SINQ at PSI guarantees a highly efficient pion and muon production. The full proton beam intensity can be used while about 70\% of the beam continues with only a small energy loss to the spallation target of SINQ. Most of the recent particle physics experiments used low momentum beams of both polarities ($\mu^+$, $\mu^-$) 
for measurements with stopped muons in low mass targets. Prime examples of such efforts yielded (i) the new limit of $4.2\times10^{-13}$ (90\% C.L.) for the charged lepton flavor violating decay $\mu^+ \rightarrow {\rm e}^+ \gamma$ by the MEG collaboration,\cite{MEG16} using $3\times 10^7$/s $\mu^+$ of about 28\,MeV/c (4\,MeV kinetic energy) stopping after some degrader in an about 200\,$\mu$m thick polyethylene target and (ii) the measurements of 2S-2P Lamb shift transitions in muonic hydrogen and deuterium by the CREMA collaboration,\cite{Ant13,Poh16} stopping about 600/s $\mu^-$ of  only 3\,keV kinetic energy in a 20\,cm long target filled with about 1\,hPa of H$_2$ or D$_2$, respectively, and determining the nuclear rms charge radii of both proton and deuteron to about $4\times 10^{-4}$ precision.
These two experiments were conducted at HIPA's $\pi$E5 beamline (see the beamline overview map\cite{Hallenplan}) which presently is the world's highest intensity low momentum muon beam for particle physics. Even slightly higher intensities are obtained at the $\mu$E4 beam\cite{Hallenplan,Pro08} which is usually dedicated to condensed matter studies with muon spin rotation ($\mu$SR) but has recently also been applied to study Mu production into vacuum.\cite{Ant12} A very successful series of experiments was conducted by the MuLan,\cite{MuLan} MuCap,\cite{MuCap} and MuSun\cite{MuSun} collaborations, initially in the $\pi$E3 and later in the $\pi$E1 beam areas.\cite{Hallenplan} The 1\,ppm measurement of the positive muon lifetime by MuLan resulted in a 0.5\,ppm determination of the Fermi coupling constant $G_{\rm F}$. MuCap measured the $\mu^-$ lifetime in a high-purity protium target and determined the singlet capture rate to 1\% and by that the weak pseudoscalar coupling $g_{\rm P}$ of the proton to 7\%. MuSun is under way to determine the muon capture rate on deuterons by the same technique. 

Various new muon experiments are under way and progress has recently been reported at PSI's annual particle physics users meeting.\cite{BVR2016}
The MEG II experiment aims at a sensitivity to $\mu^+ \rightarrow {\rm e}^+ \gamma$ of $4 \times 10^{-14}$.  The Mu3e collaboration pursues a search for the charged lepton flavor violating decay $\mu^+ \rightarrow {\rm e}^+ {\rm e}^+ {\rm e}^-$ in two phases, first aiming at $10^{-15}$ and later $10^{-16}$ at the new HiMB.
Studies for HiMB already benefit conventional meson target designs.\cite{Ber16} HiMB will provide $10^{10} \mu^+$/s below 30\,MeV/c to the phase-2 Mu3e experiment. The beamline will  
open up new possibilites for statistically limited muon experiments in general. The MUSE collaboration will compare muon and electron scattering on hydrogen at the $\pi$M1 beamline\cite{Hallenplan} and decisively test the difference in rms charge radii obtained so far by electron scattering and ordinary hydrogen spectroscopy versus the muonic atom result.\cite{Ant13}
The CREMA collaboration has meanwhile finished data taking on the 2S-2P Lamb shift transitions in $^{3,4}$He and is now turning to the laser spectroscopy of the ground state hyperfine splitting in H and $^3$He. Also spectroscopy of heavy muonic atoms is regaining interest, e.g.,
the MuX collaboration is setting out to measure the charge radii of $^{226}$Ra and other heavy, radioactive isotopes.
A lot of progress is also being made by the muCool collaboration with the development of a very high brightness beam of slow $\mu^+$.\cite{Taq06,Bao14,BVR2016,CPT16_muCool} Together with Ref.\ \refcite{Ant12} this development will allow for an improved measurement of the Mu 1S-2S transition\cite{CPT16_Mu1S2S} and
pave the way for a test of the free fall of Mu,\cite{CPT16_Mu-grav} the mass of which is dominated by the antimatter, second generation lepton. 

\section*{Acknowledgments}
Illuminating discussions with colleagues, especially at CPT'16, 
and continued support of PSI's
accelerator and beam line groups and PSI/ETH technical services,  
the Swiss National Science Foundation (200020\_159754) 
and the ETH Z\"urich (ETH-35 14-1) 
are gratefully acknowledged.

\end{document}